**Enhanced Superconductivity in Multilayer FeSe Films by Simplified Molecular Beam Epitaxy**

*Maria Hilse*, Hemian Yi, Zhe Chen, Jessica L. Thompson, Kalana D. Halanayake, Danielle Reifsnyder Hickey, Seong H. Kim, Cui-Zu Chang, Nitin Samarth, and Roman Engel-Herbert*

M. Hilse
Two-Dimensional Crystal Consortium (2DCC) -Materials Innovation Platform, The Pennsylvania State University, University Park, USA
Materials Research Institute, The Pennsylvania State University, University Park, USA
Department of Materials Science and Engineering, The Pennsylvania State University, University Park, USA
E-mail: mxh752@psu.edu

H. Yi
Department of Physics, The Pennsylvania State University, University Park, USA
Tsung-Dao Lee Institute and School of Physics and Astronomy, Shanghai Jiao Tong University, Shanghai, China

Z. Chen
Materials Research Institute, The Pennsylvania State University, University Park, USA
Department of Chemical Engineering, The Pennsylvania State University, University Park, USA
State Key Lab of Fluid Power and Mechatronic Systems, Zhejiang University, Hangzhou, China

J. L. Thompson
Department of Chemistry, The Pennsylvania State University, University Park, USA

K. D. Halanayake
Department of Chemistry, The Pennsylvania State University, University Park, USA

D. Reifsnyder Hickey
Materials Research Institute, The Pennsylvania State University, University Park, USA

Department of Materials Science and Engineering, The Pennsylvania State University, University Park, USA
Department of Chemistry, The Pennsylvania State University, University Park, USA

S. H. Kim
Materials Research Institute, The Pennsylvania State University, University Park, USA
Department of Chemical Engineering, The Pennsylvania State University, University Park, USA

Cui-Zu Chang
Department of Physics, The Pennsylvania State University, University Park, USA

N. Samarth
Department of Physics, The Pennsylvania State University, University Park, USA
Department of Materials Science and Engineering, The Pennsylvania State University, University Park, USA
Q-NEXT, Argonne National Lab, Lemont, IL

R. Engel-Herbert
Paul-Drude-Institut für Festkörperelektronik, Leibniz-Institut im Forschungsverbund Berlin e.V., Berlin, Germany

Funding:
NSF Cooperative Agreement No. DMR- 2039351
Penn State University (PSU) Materials Research Science and Engineering Center (MRSEC) for Nanoscale Science (DMR-2011839)
Gordon and Betty Moore Foundation's EPiQS Initiative (GBMF9063 to C. -Z. C)
Penn State Eberly College of Science, Department of Chemistry, College of Earth and Mineral Sciences, Department of Materials Science and Engineering, and Materials Research Institute start-up funds

Abstract:

Multi-unit-cell (UC) β-FeSe films grown on $SrTiO_3(100)$ continue to attract attention because of the significant enhancement in the superconducting transition temperature ($T_c$) compared to that in bulk FeSe. In prior reports of molecular beam epitaxy (MBE)-grown β-FeSe/$SrTiO_3(100)$, elaborate growth protocols have been used to achieve enhanced $T_c$, leading to a general belief that careful pre-treatment of the $SrTiO_3$ substrate and post-growth annealing in ultrahigh vacuum (UHV) are essential. Here, we report a greatly simplified protocol for the MBE growth of superconducting multi-UC β-FeSe films on $SrTiO_3(100)$, eliminating the need for careful substrate pre-treatment and post-growth UHV annealing while still achieving an enhanced $T_c$. With appropriate capping, epitaxial films with 14 UC thickness exhibit a zero-resistance transition temperature $T_c \sim 20$ K in *ex situ* electrical transport measurements. The MBE optimization process is guided by the growth-parameter dependencies of film morphology and structural properties, as characterized by reflection high-energy electron diffraction, X-ray diffraction, atomic force microscopy, and scanning transmission electron microscopy.

1. Introduction

Among superconducting iron-based compounds, FeSe, a bulk superconductor with a transition temperature $T_c \sim 8$ K[1], has attracted much attention for more than a decade after the discovery of a striking enhancement in $T_c$ when a single unit cell (UC) layer of FeSe is grown on $SrTiO_3$(100) by molecular beam epitaxy (MBE).[2–6] In such single UC FeSe films, scanning tunneling microscopy (STM) and angle-resolved photoemission spectroscopy (ARPES) show that a superconducting gap opens at temperatures below 65 K, far above the bulk $T_c$.[2,7–20] This gap opening is accompanied by a downturn in the sample sheet resistance, which eventually becomes zero at a somewhat lower temperature ($T_{c,0}$ <~ 30 K). Although the microscopic mechanism responsible for the enhanced $T_c$ is not entirely understood, there is consensus that it likely arises from charge transfer from the substrate into the ultra-thin FeSe film, accompanied by enhanced electron-phonon coupling across the interface[19,21,22] and possibly the quenching of spin density waves in the interfacial FeSe region.[23] The superconducting transition temperature as measured by *ex situ* electrical transport can vary significantly with parameters such as film thickness and post-growth annealing,[20] the choice of capping layer,[24] and structural disorder.[25]

The nature of superconductivity in multi-UC FeSe films on $SrTiO_3$(100) presents an unresolved puzzle of particular interest in this context. STM studies of films of thickness 2 UC and greater do not show any superconducting gap,[2,10,11,15] but electrical transport in multi-UC films shows a $T_c$ that decreases with increasing film thickness.[20] This implies that superconductivity is always confined to a region near the interface with the $SrTiO_3$ substrate and that the upper layers of a multi-UC FeSe film do not exhibit any proximity-induced superconductivity, as observed in FeTe-based heterostructures.[26] One hypothesis that accounts for this behavior in multi-UC FeSe is that the net charge transfer from the substrate is redistributed along the growth axis, thus lowering excess interfacial charge responsible for an enhanced $T_C$ at the interface.[20] This charge redistribution allows spin density waves to persist in regions of the film away from the interface, thus preventing the formation of Cooper pairs.[23] A full understanding of the thickness variation of $T_c$, however, is still lacking.

To better understand the nature of superconductivity in multi-UC $FeSe/SrTiO_3$ films, it is important to first improve our control over the underlying materials by systematically studying structure-property relationships. Although the prior literature shows many different growth protocols that produce $FeSe/SrTiO_3$ films with enhanced $T_c$, a few sample growth and heterostructure preparation steps are believed to be essential, specifically:

(a) An ultra-clean substrate surface of a specific $TiO_2$ termination needs to be realized across the wafer using *ex situ* chemical and thermal treatment, followed by an additional *in situ* thermal substrate preparation before growth.[14,22–24,27–31] However, we find that it is difficult to reproduce the effects of thermal-chemical wafer preparation recipes, at least as judged by surface morphology. Even wafers from the same supplier and prepared in the same batch differ in surface morphology (Figure **S1(a)**, **(b)**).

(b) The FeSe film thickness needs to be restricted to a few UC. Although FeSe films of 5 UC to 50 UC on $SrTiO_3$ show an enhanced $T_c$ compared with bulk, this enhancement decreases with increasing thickness, and achieving superconductivity in these films required extraordinary post-growth annealing times (~35 hours) for optimal results.[20,22]

(c) Post-growth film annealing in ultra-high vacuum (UHV) for many hours is essential for creating a Se vacancy concentration of about (1–5)%,[36,38,41,47] which is accompanied by electron doping of the FeSe layer.[2,14,20–24,27–30,32,48] We note, however, that FeSe growth on anatase $TiO_2$ showed an enhanced $T_c$ without any post-growth UHV annealing.[31]

In this work, we present a systematic study of the growth of relatively thick (14 UC) FeSe films on mixed-termination $SrTiO_3(100)$ substrates. By optimizing the substrate temperature and the Se/Fe flux ratio (FR) *in situ*, we develop a highly simplified growth protocol that reproducibly yields multilayer FeSe films with enhanced superconductivity, reaching $T_c$ values up to 20 K. This approach does not require either pre-growth substrate preparation or post-growth UHV annealing. Our results thus circumvent all three growth requirements believed to be essential for achieving enhanced $T_c$ in FeSe films grown on $SrTiO_3(100)$ and provide interesting insights into the fields of bulk and interfacial superconductivity.

## 2 Results and Discussion

### 2.1 Influence of deposition parameters on surface morphology of multilayer β-FeSe films

We grew 14-UC-thick FeSe films on $SrTiO_3(100)$ prepared to expose an atomic step terrace surface morphology with smooth edges but with mixed termination (see Figure **S2**) at temperatures of 330°C, 390°C, and 450°C and precisely controlled Se/Fe FRs of 2.5, 5, 12, and 24. The FRs were calibrated using a quartz crystal microbalance (QCM) before each film growth. Reflection high-energy electron diffraction (RHEED) was used as an *in situ* monitor of FeSe film growth. The FeSe film morphology and crystallinity of uncapped films were assessed *ex situ* using AFM and XRD within 20 min of removing the samples from UHV, and vacuum-packaged samples were used to transport them to the respective characterization tools to minimize oxidation of the films. (A time window of 2 h was previously determined during which the intrinsic properties of thin FeSe flakes in air at ambient pressure conditions were conserved. [49])

Pure β-FeSe phase formation was confirmed for all samples by XRD, and the observed RHEED patterns showed streaky, intense FeSe diffractions with defined Kikuchi lines indicative of smooth film growth (Figure S3). However, as shown in **Figure 1**, the films' morphology displayed an unexpectedly strong dependence on the growth temperature and FR. Atomically smooth and continuous films with a root mean square (rms) roughness of only 1.43 nm and atomic terraces (terrace edges are represented by the narrow bright horizontal yellow lines in the inset of Figure 1) were realized at a 450°C growth temperature and a low FR of 2.5. In contrast, disconnected tall islands with smooth top surfaces were observed at lower temperatures and/or higher FRs, eventually culminating in an rms roughness of 18.89 nm for the sample grown at 330°C and FR = 24. Growth at smaller FRs than 2.5 resulted in mixed – β-FeSe/elemental Fe – phase formation that is not shown here. The irregularly shaped small islands that appear to be floating on top of the otherwise atomically smooth FeSe films and/or islands in Figure 1 were attributed to oxidation onsetting of the FeSe, as this island formation scaled with the time that the sample had been taken out of the UHV environment and was seen as well in reports on exfoliated flakes.[49]

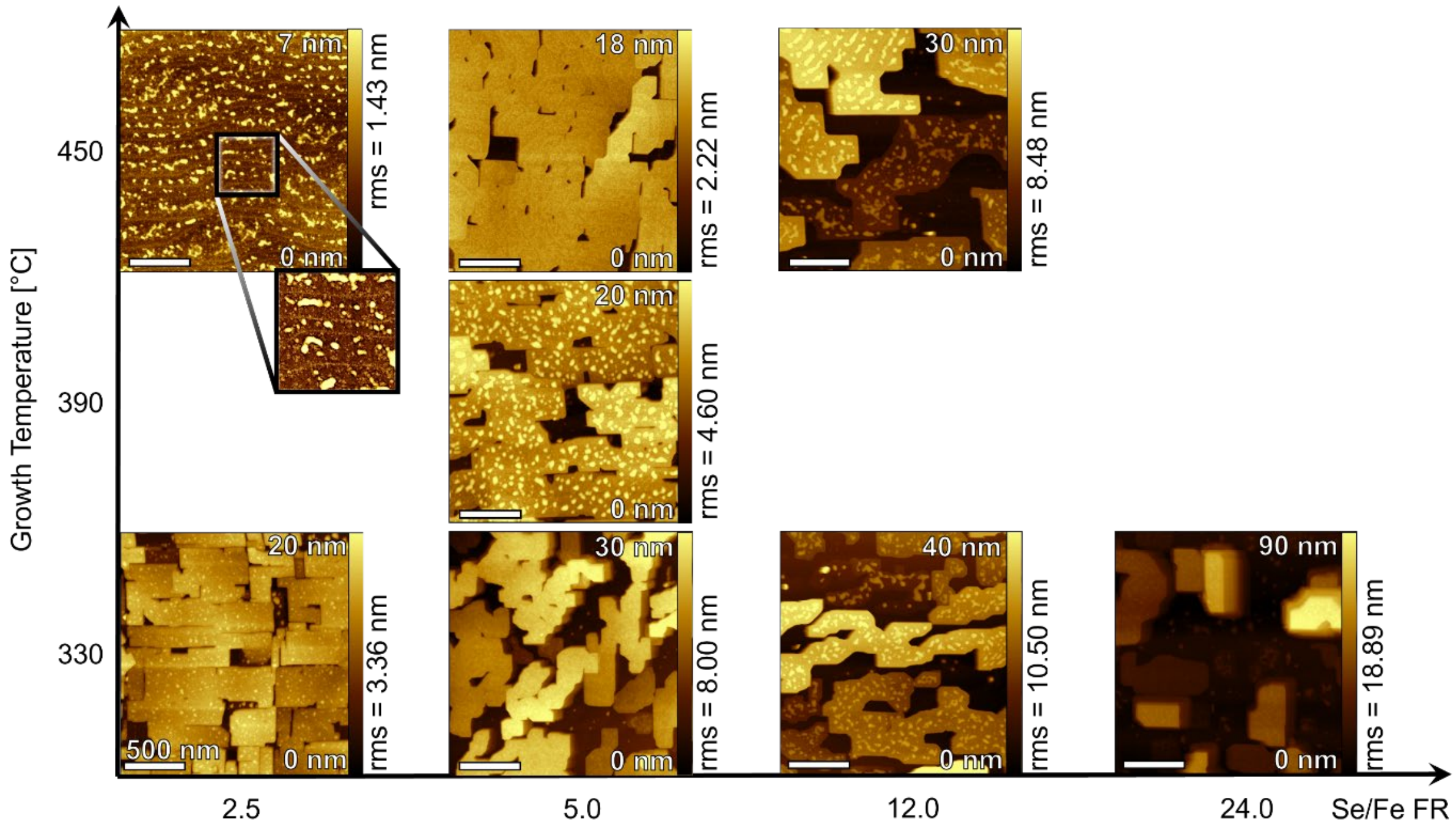

**Figure 1.** Surface morphology measured in AFM for FeSe films grown at different temperatures (lowest at the bottom to highest at the top) and varying Se/Fe FR (lowest at the left to highest at the right) with a universal indicated scale bar of 500 nm and printed rms roughness. The highlighted zoom-in region for the sample at the top left (450°C, FR = 2.5) with the lowest rms roughness shows the atomic terrace edges, i.e., the horizontal bright yellow lines observed in this FeSe film.

Figure 1 thus demonstrates that a high Se overpressure during FeSe growth, i.e., high FR (towards the right side of Figure 1), strongly promotes island growth, whereas high growth temperatures and low Se oversupply (top left corner of Figure 1) enable a layer-by-layer or step-flow growth mode. A similar sensitive temperature dependence of FeSe film morphology was observed in pulsed laser deposition.[50,51]

**2.2 Influence of deposition parameters on structural properties of multilayer β-FeSe films**

The RHEED patterns remained unexpectedly streaky, with well-defined Kikuchi lines and low background intensity indicative of smooth surfaces, for all samples except the one grown at 330°C with FR = 24 (see Figure **S3**). Therefore, RHEED showed limited sensitivity to variations in surface morphology.[52] Noticeable changes in RHEED were observed only for films with rms roughness exceeding 18 nm. The observed insensitivity of RHEED to the FeSe surface morphology is most likely due to the very peculiar shape of the formed FeSe islands. As they present pillars with irregular lateral circumference but emerge from the $SrTiO_3$ surface in a close-to-ideal step function with nearly the same average height and display an atomically smooth top island surface, they nearly present a continuous atomically flat film in RHEED since the probing depth in RHEED is limited to the topmost atomic layers on the sample. A more in-depth analysis of electron diffraction is required to elucidate the limits of RHEED under such circumstances, which is not the scope of this study.

Because AFM directly probes film morphology, we used rms roughness to distinguish between smooth layer growth and predominantly island growth. **Figure 2**(a) plots the rms roughness over the growth temperature and the Se/Fe FR and clearly demonstrates the inverse scaling of rms roughness with increasing growth temperature on the left blue scale (with fixed Se/Fe FR = 5) and the nearly linear scaling of rms roughness with increasing FR on the right pink scale for the two sets of data at 330°C and 450°C fixed growth temperature as discussed earlier. The optimum region for smooth, continuous FeSe films and low rms roughness values is illustrated by the blue top-left corner, towards higher growth temperatures, and by the pink lower line of low Se/Fe FRs.

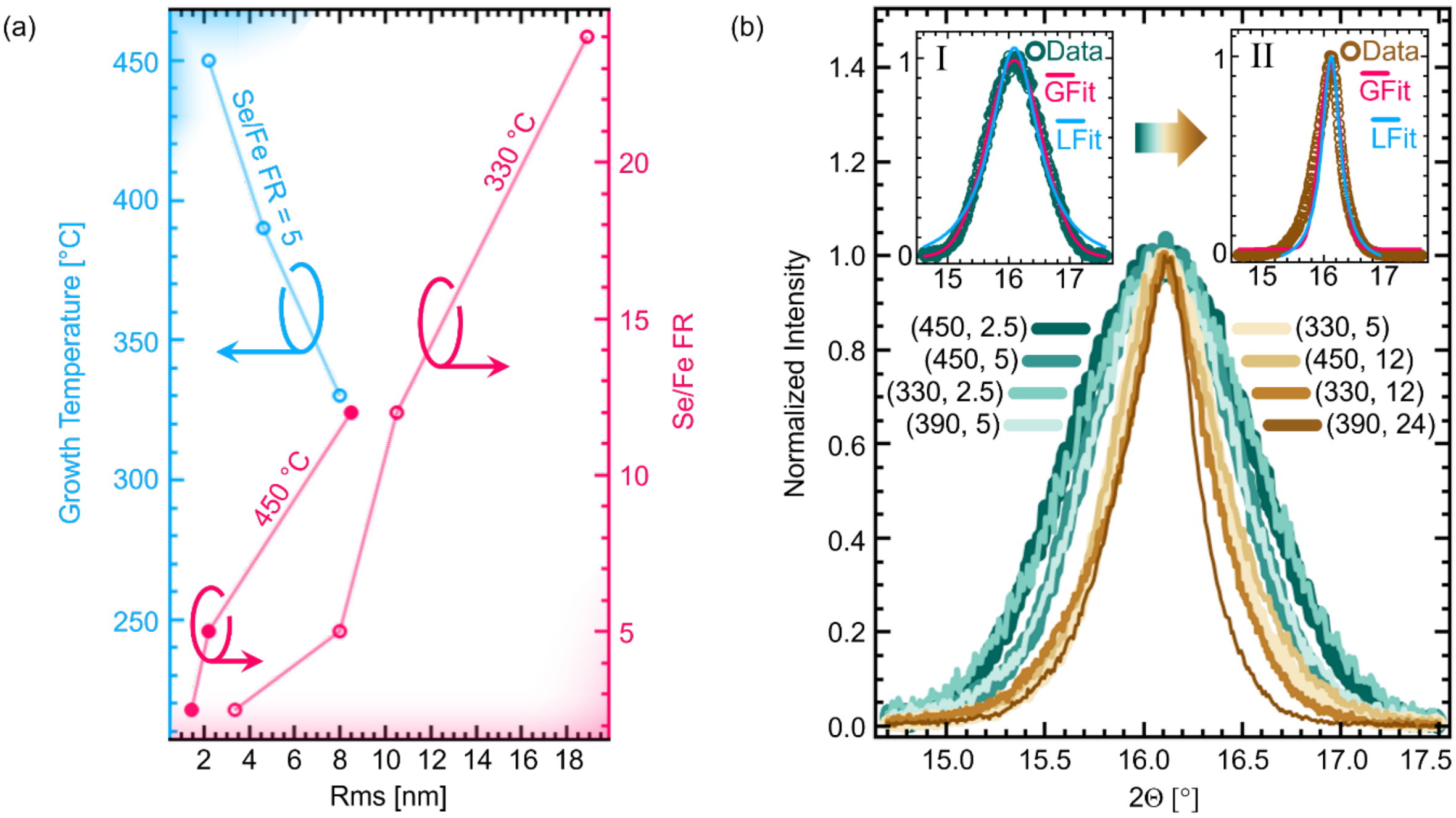

**Figure 2.** (a) The rms values obtained for all samples shown in Figure 1 plotted over the FeSe film growth temperature and Se/Fe FR. (b) XRD FeSe(001) diffraction peaks measured along 2Θ for all samples of Figure 1 sorted by rms roughness reflected through the color scale from green to brown (lowest rms value = curve in dark green, highest rms value = curve in dark brown). Insets I and II show the same FeSe(001) diffraction for only the sample with the lowest

(450°C, FR = 2.5) and highest (330°C, FR = 24) rms value, respectively, with respective fits of a Gaussian (pink) and a Lorentzian (blue) distribution function.

XRD confirmed the formation of the β-FeSe phase for all samples shown in Figure 1 (Figure **S3**). This figure also shows that the diffraction peak shape of all detected FeSe-related peaks in 2Θ-ω changes with the rms roughness of the FeSe film: the morphology of the film is imprinted into the diffraction peak shape in out-of-plane XRD, with Gaussian-like peak shapes for smooth and continuous films and Lorentzian-like ones for pronounced island growth. To the best of our knowledge, this is the first report of a transition in diffraction peak shape from Gaussian to Lorentzian with rms roughness in XRD 2*Θ*-ω-scans. The well-established theory of FWHM broadening in the 2Θ-ω XRD geometry relates the peak width to a decrease in crystalline domain sizes and/or an increase in non-uniform strain, i.e., dislocations in the film.[53–55] The FWHM values of the FeSe (001) diffraction peak for the samples in Figure 1 are summarized in Table **S1** in the Supporting Information document, and range from (1.075 ± 0.003)° for the sample with the lowest rms roughness grown at 450°C and a FR of 2.5 to (0.468 ± 0.005)° for the sample with the highest rms roughness grown at 330°C and a FR of 24. Assuming a perfectly single-crystalline layer without the presence of dislocations or microstrain, the Scherrer equation relates the FWHM measured in the 2Θ-ω XRD geometry to the film thickness, yielding a film thickness of (7.63 ± 0.05) nm for the sample with the lowest rms roughness. This value is in good agreement with a nominal deposited film thickness of 7.70 nm for 14 UC FeSe, demonstrating that a non-uniform residual strain component in the films is highly unlikely. We do not observe any peak shift in the FeSe-related diffraction peaks in the presented XRD analysis either, ruling out uniform residual epitaxial strain in the FeSe films. For samples with rms roughness greater than 8 nm, the FWHM values decreased with increasing rms roughness. This might be due to the increase in island heights triggered by lower growth temperatures and higher Se oversupply, which could appear as an effectively thicker, though more porous film over the macroscopic scales in XRD.

### 2.3 Influence of deposition parameters on stoichiometry of multilayer β-FeSe films

Extracting the diffraction peak intensities of the FeSe-related (001) and (002) peaks in 2Θ-ω-XRD-scans for all samples presented in Figure 1 allows for identifying the stoichiometry in the β-$Fe_{2-x}Se_x$ via structure factor analysis. All FeSe peak intensity ratios for the respective sample growth conditions from Figure 1 are given in Table 1. Structural calculations assuming room temperature β-$Fe_{2-x}Se_x$ allow for XRD structure factor calculations, which served as comparison to the experimentally extracted (001)/(002) peak intensity ratios to obtain the Se content x printed in Table 1.[56]

**Table 1.** Summary of growth conditions (growth temperature $T_G$ and Se/Fe FR), surface rms roughness measured by AFM, FeSe peak intensity ratio $I_{001/002}$ measured in 2θ-ω XRD geometry, and calculated Se content x in the β-$Fe_{2-x}Se_x$ thin film for all samples contained in Figure 1, Figure 2, and Figure S3, sorted by rms roughness from low to high.

| $T_G$ [°C] | Se/Fe FR | Rms roughness [nm] | $I_{001/002}$ | Se content x |
|---|---|---|---|---|
| 450 | 2.5 | 1.43 | 44 ± 15 | 1.00 ± 0.04 |
| 450 | 5 | 2.22 | 26 ± 3 | 1.06 ± 0.02 |
| 330 | 2.5 | 3.36 | 20 ± 8 | 1.12 ± 0.07 |
| 390 | 5 | 4.60 | 31 ± 7 | 1.03 ± 0.02 |
| 330 | 5 | 8.00 | 23 ± 2 | 1.08 ± 0.02 |
| 450 | 12 | 8.48 | 33 ± 4 | 1.03 ± 0.01 |
| 330 | 12 | 10.50 | 21.4 ± 0.9 | 1.11 ± 0.01 |
| 330 | 24 | 18.90 | 20 ± 1 | 1.118 ± 0.007 |

Although there is no clear trend in Se content with varying the FR at a fixed growth temperature, growth temperatures of 390°C and higher consistently show lower Se levels, with values scattered slightly above 1. All samples grown at 330°C exhibit excess Se of 8-12%. Within the error margins, only one set of growth conditions (450°C and FR of 2.5) falls within or is close to the (001)/(002) XRD peak intensity ratio window of 45 to 115, which was previously identified as resulting in superconducting FeSe thin films.[56] Accordingly, these growth conditions have the potential to stabilize slightly Fe-rich film compositions (x = 1.00 ± 0.04) that satisfy the stoichiometric requirement for superconducting properties in FeSe. Although more in-depth growth kinetics studies for the epitaxy of FeSe in MBE are needed to elucidate the mechanism and quantitatively relate stoichiometric changes with growth conditions, this study identified optimal growth conditions for β-FeSe thin films.

### 2.4 Microstructure and electronic properties of optimized multilayer β-FeSe films

The FeSe film grown under optimized conditions (450°C and FR = 2.5) showed a sheet resistance of 929 Ω/□, and a sheet (hole) carrier concentration of $7.48 \times 10^{15}$ $cm^{-2}$ in room-temperature Hall transport measurements (van der Pauw geometry) across the $(1 \times 1)$ $cm^2$ large sample. Although reference values for the metal FeSe at room temperature are scarce, they appear to fall within the literature range.[1,20,39,57–59] Two additional samples were prepared under optimized growth conditions and capped with FeTe/Te to conduct STEM characterization and probe the superconducting properties of the optimized films *ex situ*. For those samples, the $SrTiO_3$ wafer surface preparation recipe, FeSe growth (450°C and FR = 2.5), and thickness (14 UC) were kept identical to the above-described procedure. No UHV post-growth annealing step or any other post-growth treatment was performed after FeSe growth. The capping layer consisted of 14-UC-thick FeTe grown at 300°C for sample 1, and the identical FeTe layer plus an additional amorphous Te (a-Te) layer of about 15 nm deposited at room temperature in the MBE for sample 2.

**Figure 3** shows low- and high-magnification annular dark field (ADF)-STEM images of sample 1 in (a) and sample 2 in (b) together with elemental maps obtained from energy-dispersive X-ray (EDX) spectroscopy analysis. The substrate and the films of FeSe, FeTe, and a-Te are clearly visible in both samples, and exhibit high uniformity across the entire STEM sample. The nominal deposited thicknesses of 14 UC FeSe and 14 UC FeTe in sample 1, and the same plus a 12 nm a-Te layer in sample 2, were confirmed by STEM analysis in Figure 3. A dark layer with irregular and wavy edges towards the FeTe and a-Te additionally appears on the top of the heterostructures in both samples. These dark features are attributed to an oxide layer, as indicated by additional arrows in the high-magnification ADF-STEM image of sample 1 in Figure 3(a). The *ex situ* sample preparation conditions and possible focused-ion-beam (FIB) thinning damage are likely the cause of the observed oxidation layer. In sample 1, i.e., Figure 3(a), these dark circular features were found to be rich in Fe and O according to EDX mapping. Although a consistent number of layers was observed in the FeSe and FeTe films in sample 1 throughout the sample's wide range, as demonstrated by the low-magnification ADF-STEM image in Figure 3(a), a few regions with varying FeTe thicknesses and different grain boundaries in the FeTe crystal structure were also observed, as is apparent in the high-magnification ADF-STEM image in Figure 3(a). Although both samples show a dark layer approximately 2 atomic sheets thick at the $SrTiO_3$/FeSe interface in the high-magnification STEM images in Figure 3, the well-ordered atomic structure across the interface and step edges on the substrate surface were still resolved.

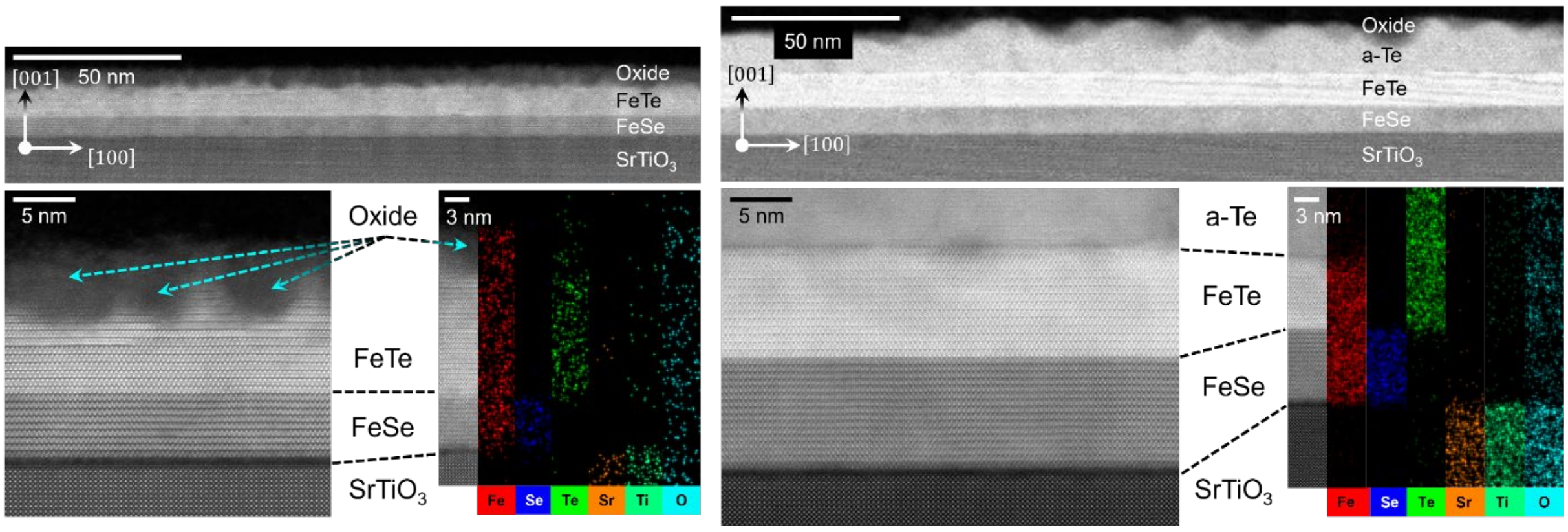

**Figure 3.** ADF-STEM images at low and high magnification and EDX elemental maps of the [010] zone axis of (a) sample 1 and (b) sample 2. The 14-UC-thick FeSe film in both samples was grown under optimized conditions at 450°C with an FR of 2.5. For sample 1, the FeSe film was capped with a 14-UC-thick FeTe layer grown at a lower temperature than the FeSe film. For sample 2, an additional protective amorphous Te (a-Te) layer of about 12 nm was deposited onto the identical FeTe/FeSe heterostructure of sample 1.

In contrast to sample 1, the number of FeSe and FeTe layers in sample 2 showed remarkable consistency across the entire range of the prepared specimen. Dark circular features, i.e., oxide inclusions near the FeTe layer, were not observed in sample 2. Film damage from oxidation is entirely contained within the a-Te, demonstrating that the additional a-Te capping layer is an effective protection against oxidation. The nominal sample structure and film compositions are confirmed through STEM and EDX. A well-ordered atomic structure across the interface is also observable.

The temperature-dependent sheet resistances of FeSe films are shown in **Figure 4**(a). Both samples show low room-temperature resistance and metallic behavior upon cooling, followed by a superconducting transition below 30 K. The zero-resistance state is reached at 18 K for sample 1 and 20 K for sample 2. These values are significantly higher than the established $T_c$ for bulk FeSe ($T_c \sim 8$ K);[1] prior studies of MBE-grown FeSe/$SrTiO_3$ films of similar thickness have shown $T_c$ values either somewhat lower [22] or comparable values [Ref. 20]. We note again that both these latter studies used a more elaborate pre-growth and post-growth protocol.

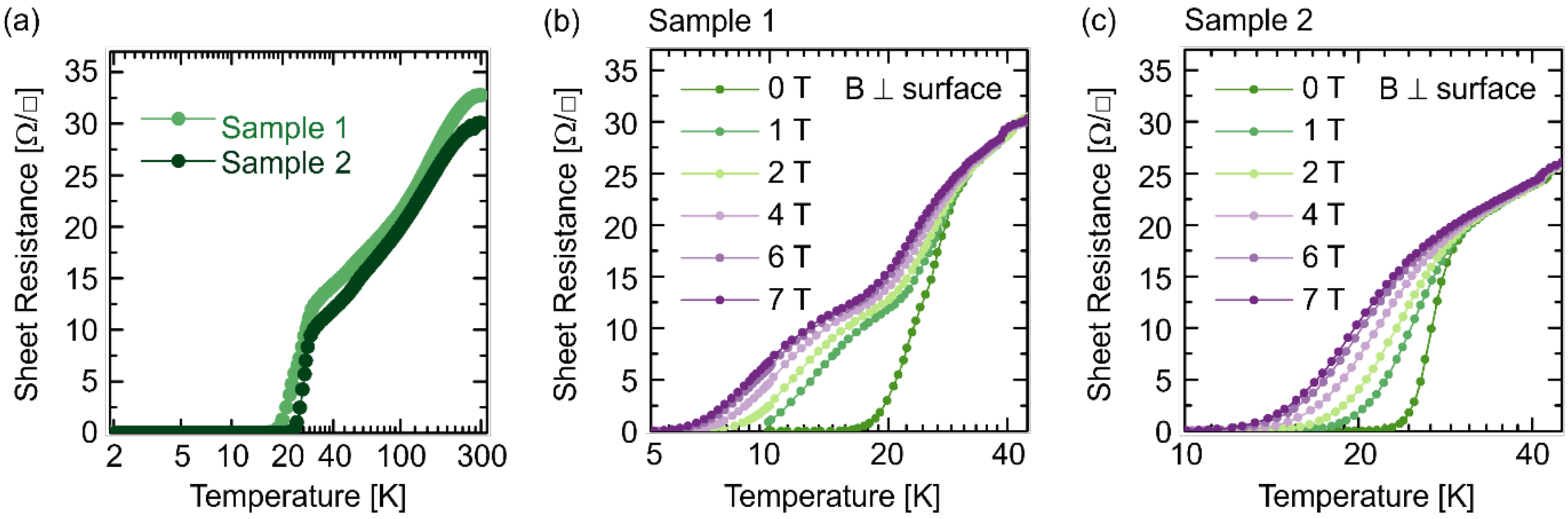

**Figure 4.** (a) Electrical transport measurement of sheet resistance versus temperature at zero external magnetic field cooling for samples 1 and 2. (b) and (c) demonstrate the same measurement carried out in different external magnetic fields applied perpendicular to the film surface for samples 1 and 2, respectively.

Figure 4(b) and (c) show the temperature-dependent resistances of samples 1 and 2 measured between 5 K to 50 K and 10 K to 50 K, respectively, under perpendicular magnetic fields from 0 T to 7 T. The superconducting onset temperatures in both samples remain nearly unchanged with the applied external field. The $T_c$, however, shifts continuously to lower temperatures with increasing external magnetic field in both samples. The lowest $T_c$ values of 6 K and 11 K were observed at the highest applied external field of 7 T for samples 1 and 2, respectively. In addition, the transition into the superconducting phase for sample 1 is two-step in the presence of an external magnetic field. The constant onset temperature and the two-step transition behavior with external field, which contrast with what has been observed for bulk β-FeSe, are indicative of a vortex motion typical of type-II superconductors.[1,50,60,61]

Overall, sample 2 with the additional a-Te cap showed lower room-temperature resistance, a higher Tc, a single-step transition, and a narrower transition width than sample 1. This might be the result of better ex situ sample preservation enabled by the additional capping layer, as evidenced by STEM characterization. It may also result from slight differences in the configuration of the starting substrate surface. XRD analysis of both samples yielded c lattice parameters of 5.49 ± 0.01 Å for FeSe and 6.1 ± 0.4 Å for FeTe. Those values agree with the reported lattice parameters of unstrained FeSe and FeTe crystals, indicating the absence of any significant epitaxial strain or Te doping resulting from potential interdiffusion of Se and Te across the FeSe/FeTe interfaces.[62,63] Nevertheless, we observe an enhanced $T_c$ in our multilayer FeSe films, well above the reported maximum values attained through pressure and Te-doping[42–46] and similar to prior observations using long post-growth annealing (Ref. 20). We note that our study is the first to observe superconductivity above 8 K in FeSe layers thicker than 2 UC solely by engineering the FeSe growth conditions.

As discussed in the introduction, the mechanism for the persistence of this enhanced superconductivity in electrical transport measurements of multilayer FeSe/STO films (albeit with decreasing $T_c$) remains unclear. The thickness dependence of $T_c$ in the FeSe/$SrTiO_3$ system was recently calculated using first-principles density functional theory (DFT), exploring the effects of electron correlation via spin-wave exchange mechanisms.[64] The DFT calculations qualitatively reproduced the established reported trend in $T_c$ with FeSe layer thickness, i.e., 1-UC-thick films show an enhanced $T_c$, while $T_c$ reverts quickly to the bulk value when the thickness is increased by only one to three more UCs without including possible phonon or doping contributions from the substrate in the model. The experimental results in Ref. 20 and the present paper thus suggest that the electron distribution and electron-phonon coupling across the substrate/film interface likely play an important role in the observed increase in $T_c$. Further studies are required to elucidate the current pathway and the effect of additional FeSe layers. It seems possible, though, that the $T_c$ might be even further enhanced through additional property-targeted FeSe layers and/or other capping materials based on the results of this study.

## 3 Conclusion

In summary, we report a highly simplified protocol for growing multilayer FeSe films on (001) $SrTiO_3$ with an enhanced $T_c$ of up to 20 K. This is achieved by optimizing growth solely in terms of surface morphology and stoichiometry through precise in situ control of the substrate temperature and the Se/Fe FR. The optimized growth protocol does not require fine-tuning of the pre-growth $SrTiO_3$ surface termination or post-growth sample stoichiometry via annealing. Our results demonstrate the potential of *in situ* control of growth parameters and underscore the importance of basic growth kinetics studies. Such studies provide further understanding of how growth parameters influence FeSe film morphology, crystallinity, and stoichiometry/doping levels and, consequently, the resulting superconducting properties.

## 4 Experimental Section

*MBE:* Substrate preparation was carried out as described in the Supporting Information. After loading the substrates into the UHV system, a 1 h heating cycle to 120°C was performed in the load lock to remove residual water films from the wafers. A model R450 MBE reactor from DCA Instruments (DOI: 10.60551/gqq8-yj90) was used for FeSe, FeTe, and Te depositions, equipped with individual Knudsen effusion cells that evaporated ultra-pure solid source Fe, Se, and Te charges. The background pressure in the MBE reactor was kept below 8 × $10^{-10}$ Torr during deposition.

*QCM:* Before film depositions, Fe, Se, and Te fluxes were calibrated using a QCM model Eon from Colnatec, which was inserted into the reactor at the sample position. Physical film thickness measurements by XRD were used to find tooling factors for flux calibration. Typical fluxes for Fe were on the order of 1.5 × $10^{13}$ $cm^{-2}s^{-1}$, and Se and Te fluxes ranged 2.5 to 24 times higher than Fe. FeTe growth was carried out at 300°C and a Te/Fe FR of 2.5.

*RHEED:* Film deposition was monitored *in situ* using a RHEED system equipped with an electron gun from STAIB Instruments operated at 15 kV and a kSA 400 imaging and processing camera and software package.

*IR camera:* Substrate temperature was measured with high precision at the wafer surface using an Optris Xi80 IR camera system. The offset between the measured temperature in the IR camera and the temperature set point used for the substrate heater control was found to be -35 K, -30 K, 15 K, 30 K, and 105 K for temperatures in the IR camera of 300°C, 330°C, 390°C, 450°C, and 600°C, respectively.

*XRD: Ex situ* structural characterization was carried out using a Malvern Panalytical X'Pert$^3$ MRD 4-circle diffractometer equipped with a PIXcel 3D detector. The XRD setup was operated in high-resolution configuration using Cu $K_{\alpha1}$ radiation and a 4-bounce Ge(220) crystal monochromator with a 10 mm mask.

*AFM:* Substrate and film surface morphology was analyzed *ex situ* with a Dimension Icon Bruker AFM operated in air using ScanAsyst mode, in which a ScanAsyst-Air tip with a spring constant of 0.4 N/m was scanning the surface in PeakForce Tapping mode.

*LFM:* Friction force measurement was performed with a Bruker Multimode AFM using contact mode in ambient conditions. A Bruker CONTV Si tip (nominal spring constant: 0.2 N/m, nominal tip radius: 8 nm) was used to slide against the $SrTiO_3$(100) substrate. The applied normal load between the Si tip and the substrate was 84 nN. The AFM tip's reciprocating frequency was 1 Hz, the scan size was 1 μm, and the tip sliding speed was 2 μm/s. The normal spring constant of the AFM probe cantilever was calibrated following Sader's method.[65] The lateral sensitivity of the cantilever was calculated by comparing the measured lateral signal (in mV) on a reference sample with a known friction coefficient.[66]

*FIB and STEM:* The wafers were coated with ~15 nm of amorphous carbon using a Leica EM ACE200 sputter coater before TEM sample preparation. An FEI Scios 2 dual-beam focused ion beam (FIB)/scanning electron microscope (SEM) was then used to prepare electron-transparent lamellae for TEM analysis, using SEM accelerating voltages/currents ranging from 2– 5 kV/0.40 nA– 26 nA and ion beam accelerating voltages/currents ranging from 2– 30 kV/30 pA– 7 nA. ADF-STEM images and EDX maps were collected on a dual spherical aberration-corrected FEI Titan$^3$ G2 60-300 S/TEM, at an accelerating voltage of 300 kV, convergence angle of 25.2 mrad, and ADF collection angles of 42 – 244 mrad.

*PPMS:* Hall measurements at room temperature were performed *ex situ* within one hour of taking uncapped FeSe samples out of the UHV conditions in van der Pauw geometry on the 1 × 1 cm-sized samples using pressed-on In dots as contacts for the model Ecopia HMS-3000 probe station with an external magnetic field of 0.545 T and a current of 50 μA (two samples with lowest rms roughness), and 100 nA (sample with rms = 3.36 nm).

Temperature-dependent resistance measurements were performed *ex situ* on capped samples. Four indium dots were pressed onto the sample edge centers according to the van der Pauw geometry. Gold wires were used to connect the indium contacts on the sample to the contacts on the PPMS puck (Quantum Design Inc.). The temperature dependence of the resistances was measured during the cooling process from 300 to 2 K with an applied current of 1 μA.

Acknowledgements

This work is primarily supported by The Pennsylvania State University Two-Dimensional Crystal Consortium - Materials Innovation Platform (2DCC-MIP) funded by NSF cooperative Agreement No. DMR- 2039351. The electrical transport measurements are partially supported by the Penn State MRSEC for Nanoscale Science (DMR-2011839). J.L.T., K.D.H., and D.R.H acknowledge support from the Penn State University (PSU) Materials Research Science and Engineering Center (MRSEC) for Nanoscale Science (DMR-2011839) and start-up funds from the Penn State Eberly College of Science, Department of Chemistry, College of Earth and Mineral Sciences, Department of Materials Science and Engineering, and Materials Research Institute. The authors acknowledge use of the Pennsylvania State University Materials Characterization Lab Core Facility, Materials Research Institute, University Park, PA (RRID: SCR_012386). C.-Z.C. acknowledges the support from the Gordon and Betty Moore Foundation's EPiQS Initiative (GBMF9063 to C. -Z. C).

**Data Availability Statement**

The data that support the findings of this study are available at the following link during the peer-review process: https://data.2dccmip.org/CoC47JDk2B7E and will be made publicly available on ScholarSphere with a dedicated DOI upon acceptance for publication.

**Conflict of Interest**

The authors declare no conflict of interest.

**Author Contributions**

Maria Hilse: Substrate preparation, MBE growth, XRD and AFM characterization, Study lead, main manuscript preparation & editing

Hemian Yi & Cui-Zu Chang: Electrical transport measurements, manuscript editing

Zhe Chen & Seong H. Kim: LFM measurements, manuscript editing

Jessica L. Thompson, Kalana D. Halanayake & Danielle Reifsnyder Hickey: STEM characterization, manuscript editing

Nitin Samarth & Roman Engel-Herbert: Funding acquisition, manuscript editing

**Supporting Information**

Supporting Information containing RHEED and LFM data is available in a second file from the author.

# Supporting Information

**Enhanced Superconductivity in Multilayer FeSe Films by Simplified Molecular Beam Epitaxy**

*Maria Hilse,[1-3,*] Hemian Yi,[4,5] Zhe Chen,[2,6,7] Jessica L. Thompson,[8] Kalana D. Halanayake,[8] Danielle Reifsnyder Hickey,[2,3,8] Seong H. Kim,[2,6] Cui-Zu Chang,[4] Nitin Samarth,[3,4,9] and Roman Engel-Herbert[10]*

[1] Two-Dimensional Crystal Consortium (2DCC) -Materials Innovation Platform, The Pennsylvania State University, University Park, USA

[2] Materials Research Institute, The Pennsylvania State University, University Park, USA

[3] Department of Materials Science and Engineering, The Pennsylvania State University, University Park, USA

[4] Department of Physics, The Pennsylvania State University, University Park, USA

[5] Tsung-Dao Lee Institute and School of Physics and Astronomy, Shanghai Jiao Tong University, Shanghai, China

[6] Department of Chemical Engineering, The Pennsylvania State University, University Park, USA

[7] State Key Lab of Fluid Power and Mechatronic Systems, Zhejiang University, Hangzhou, China

[8] Department of Chemistry, The Pennsylvania State University, University Park, USA

[9] Q-NEXT, Argonne National Lab, Lemont, IL

[10] Paul-Drude-Institut für Festkörperelektronik, Leibniz-Institut im Forschungsverbund Berlin e.V., Berlin, Germany

[*] Corresponding author: mxh752@psu.edu

## Substrate Preparation and Starting Surface Morphology

$SrTiO_3$(001) substrates – crystals made in Japan – purchased from MTI Corp. were cleaned with sequential ultrasonic baths in the solvents acetone, isopropyl alcohol, and DI-water, followed by blowing off with dry nitrogen gas. Subsequently, the wafers were placed in an alumina crucible into a box furnace where they were annealed in air to 980°C for 135 / 150 min. The surface morphology of the $SrTiO_3$ wafers measured by atomic force microscopy (AFM) after the furnace anneal with a dwell time of 135 and 150 min is shown in **Figure S1**(a) and (b), respectively. As proven by AFM, the so prepared $SrTiO_3$ wafers exhibit clean terraces with atomic steps and meandering but smooth terrace edges causing very small values for the root mean square (rms) roughness. A high degree of cleanliness and smooth atomic terraces on the $SrTiO_3$ surface was found to be crucial for obtaining good wetting behavior of Fe and Se and formation of the tetragonal β-FeSe phase.[1] Differences in the degree of terrace edge meandering within one batch of $SrTiO_3$ wafers, i.e., seen across the 6 wafers in Figure S1(a) might be caused by a random variation in wafer miscut (orientation and amount) where very small miscuts [samples with large terrace widths in Figure S1(a)] along the ⟨100⟩ directions running parallel to the AFM image sides in Figure S1(a) cause more meandering of terrace edges than larger miscuts [samples with small terrace widths in Figure S1(a)] along the AFM image diagonal, i.e., ⟨110⟩ directions.
Furnace annealed $SrTiO_3(001)$ wafers were subsequently loaded into the molecular beam epitaxy (MBE) system and treated with a second anneal in ultra-high vacuum (UHV) in the loading chamber to 120°C for 1 h to remove remaining water films. Thereafter, the surfaces presented with a (1 × 1) reconstruction as shown in Figure S1(c) in reflection high energy electron diffraction (RHEED) upon transferring to the growth chamber. Highly intense first order diffraction spots, higher order diffractions of lower intensity and clearly defined Kikuchi lines indicated a clean surface. Finally, the $SrTiO_3$ surface was further conditioned by a second UHV anneal to 600°C [which was reached in the infrared (IR) camera at a thermocouple set point of about 705°C] for 2 h right before growth as the observed RHEED after the second UHV anneal, shown in Figure S1(d) has noticeably less intense background, i.e., less diffusely scattered electrons, more streaky first order diffractions, more intense higher order diffractions and more defined Kikuchi lines. Commonly surfaces reconstructed in a (3 × 2) termination after the final UHV anneal corresponding to what has been observed for stoichiometric $SrTiO_3(001)$ with at most 60% $TiO_2$ and thus mixed $TiO_2$/SrO surface coverage.[2,3]
The mixed termination of the $SrTiO_3(001)$ surface was further corroborated by lateral force microscopy (LFM) shown in **Figure S2**. The surface morphology is reflected in the LFM height channel in Figure S2 on the left depicting the typically observed atomic step terrace structure from a substrate prepared by solvent cleaning and furnace annealing as described above similar to Figure S1(a) and (b). In the lateral (friction) force map, i.e., Figure S2 middle, distinct areas of bright color in the trace map indicating high surface friction are observed. The exact same pattern with reversed contrast is revealed in the retrace friction force map shown in Figure S2 on the right. The observed pattern of surface areas with low and high surface friction corresponds to alternating regions of $TiO_2$ and SrO termination on the $SrTiO_3(001)$ surface that have been reported before. Although there seem to exist contradicting reports in the literature about the absolute friction force on a SrO and a $TiO_2$ layer, the wider terraces with lower friction force are generally attributed to $TiO_2$ terminated surface regions whereas, SrO terminations are present on the narrower terraces with higher surface friction.[4–7] Evaluating the surface coverage of $TiO_2$ terminated areas seen in Figure S2, a total amount of (89 - 85)% $TiO_2$ termination was found in the presented samples.

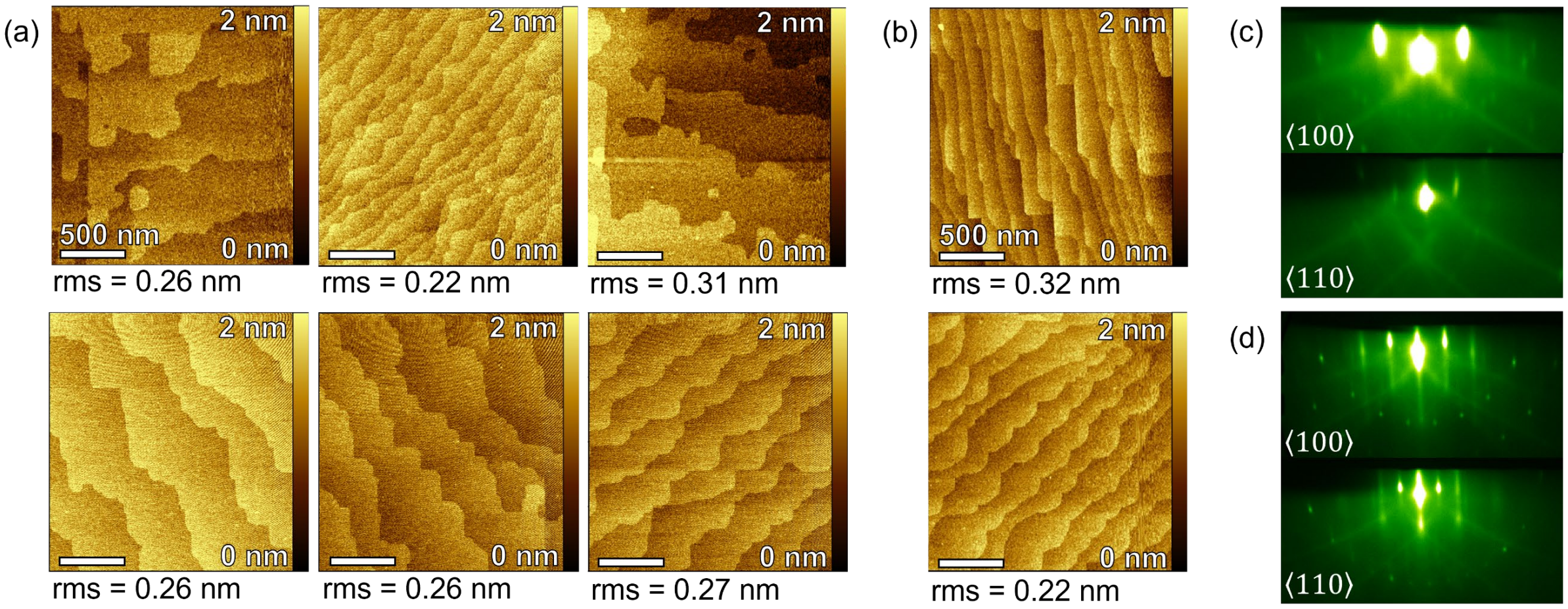

**Figure S1:** Surface morphology captured in AFM of cleaned and in a furnace annealed in air to 980°C for (a) 135 min and (b) 150 min $SrTiO_3(001)$ substrates purchased from MTI Corp. (Crystal made in Japan) before MBE growth. Exemplary RHEED patterns observed along the high symmetry axes upon loading substrates into the MBE reactor (c), and (d) after an annealing cycle in UHV for 2 h at 600°C.

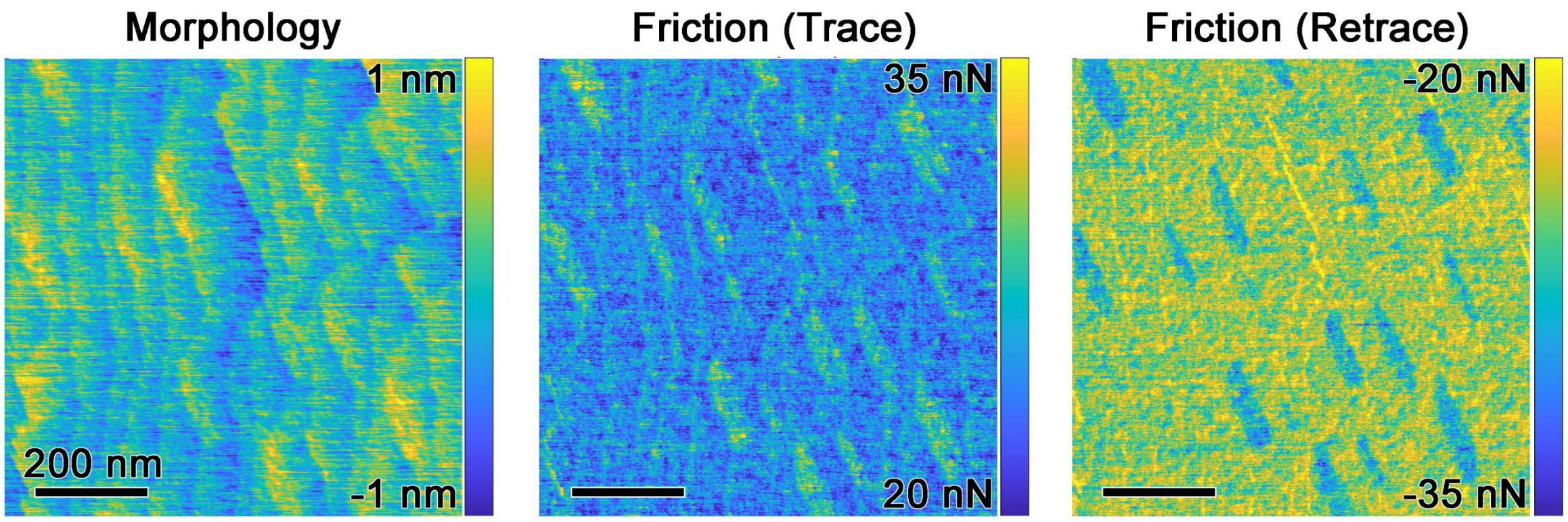

**Figure S2:** LFM images of $SrTiO_3(001)$ substrates prepared by cleaning with solvents and annealing in a box furnace in air. From left to right: surface morphology map, lateral (friction) force map - trace, lateral (friction) force map - re-trace.

**Changes observed in RHEED and XRD with Flux Ratio and Temperature**

Following the schematic of AFM data presentation in Figure 1 of the main text, Figure S3 demonstrates the observed trend in RHEED and X-ray diffraction (XRD) for FeSe film growth at different temperatures and varying Se/Fe flux ratio (FR). The position of the diffraction peaks seen in XRD correspond to the tetragonal PbO crystal structure (P4/nmm) of β-FeSe across all samples. Similarly, all samples show a high degree of crystalline ordering in the β-FeSe phase as higher order diffractions peaks up to the measured maximum of (004) were recorded.
In contrast to what is commonly seen in and thought about RHEED, there is no clear transition from streaky to spotty diffractions in RHEED that strongly correlates with the surface roughness seen in AFM. Even for the FeSe sample grown at 330°C and a Se/Fe FR of 12, which has been identified as individual, disconnected island formation of FeSe with a root mean square (rms) roughness of 10.5 nm (see Figure 1 in the main text), the RHEED pattern shows streaky intense diffractions, defined Kikuchi lines and low background intensity – all the commonly used indicators for *in situ* determination of smooth film growth. Only a small increase in diffuse background scattering intensity can be detected at low growth temperature and FRs higher or equal than 5. The RHEED pattern finally shows faint spottiness and higher diffusely scattered intensity in the background only for the sample grown at 330°C, i.e., lowest temperature and highest FR of 24, which ranges at 18.89 nm for the surface rms roughness in AFM. RHEED is thus unsuitable to evaluate FeSe film morphology *in situ* following common principles and should be handled cautiously.

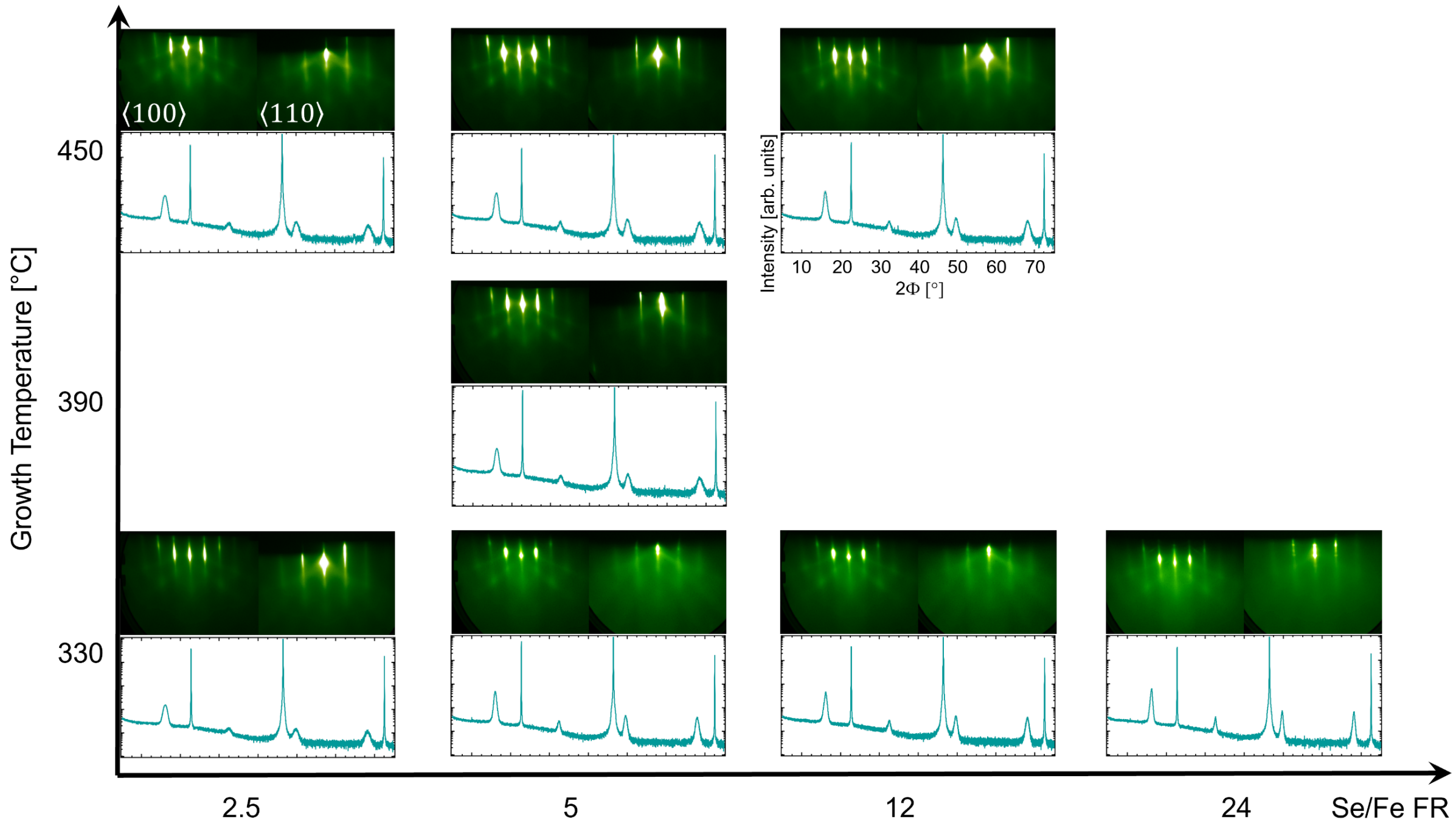

**Figure S3:** Observed RHEED patterns and *ex situ* measured XRD 2θ-ω-scans for FeSe samples grown at different temperatures (low to high temperature from bottom to top) and with varying Se/Fe FR (low to high FR from left to right).

Table S1 summarizes the growth conditions growth temperature ($T_G$), and Se/Fe FR as well as the measured rms roughness by AFM and the FWHM of the 001 diffraction peak measured in 2θ-ω XRD geometry for all samples contained in Figure 1 and Fig S3.

**Table S1:** Summary of growth conditions (growth temperature $T_G$ and Se/Fe FR) and surface rms roughness measured by AFM as well as FWHM value of the FeSe 001 peak measured in 2θ-ω XRD geometry for all samples contained in Figure 1, Figure 2, Figure S3.

| $T_G$ [°C] | Se/Fe FR | rms roughness [nm] | $FWHM_{001}$ [°] |
|---|---|---|---|
| 450 | 2.5 | 1.43 | 1.075 ± 0.003 |
| 450 | 5 | 2.22 | 0.863 ± 0.002 |
| 330 | 2.5 | 3.36 | 1.094 ± 0.004 |
| 390 | 5 | 4.60 | 0.877 ± 0.002 |
| 330 | 5 | 8.00 | 0.608 ± 0.002 |
| 450 | 12 | 8.48 | 0.676 ± 0.003 |
| 330 | 12 | 10.50 | 0.562 ± 0.003 |
| 330 | 24 | 18.90 | 0.468 ± 0.005 |